\documentclass[fleqn,twoside]{article}
\usepackage{espcrc2}

\usepackage[dvips]{graphicx}
\usepackage[figuresright]{rotating}
\usepackage{epstopdf}
\usepackage{bm}

\newcommand{\AmS}{{\protect\the\textfont2
  A\kern-.1667em\lower.5ex\hbox{M}\kern-.125emS}}

\title{\vspace{-5.0cm} 
\begin{flushright}
{\normalsize Talk presented at ``Lattice 2004'' international
symposium, June 21-26, 2004, Batavia, IL, USA}\\
\vspace{-0.2cm}
{\normalsize KEK-TH-994, CTP-MIT-3560, RBRC-473}\\
\end{flushright}
\vspace*{2.5cm}
Nucleon structure with domain wall fermions}

\author{
Shigemi Ohta\address[KEK]{Institute of Particle and Nuclear Studies, KEK,  
Tsukuba, Ibaraki 305-0801, Japan}\address[RBRC]{RIKEN BNL Research Center, Brookhaven National Laboratory, Upton, NY 11973, USA} and
Kostas Orginos\address[MIT]{Massachusetts Institute of Technology, Cambridge, MA 02139, USA}\addressmark[RBRC]
[RBCK Collaboration\thanks{Lattice 2004 talk by SO.  We thank RIKEN, Brookhaven National Laboratory and the U.S.\ DOE for providing the facilities essential for the completion of this work.}]}

\begin{document}

\begin{abstract}
We report the status of RBCK calculations on nucleon structure with quenched and dynamical domain wall fermions.  The quenched results for the moments of structure functions \(\langle x \rangle_q\), \(\langle x\rangle_{\Delta u - \Delta d}\), and \(\langle 1 \rangle_{\delta q}\) from 1.3 GeV cutoff lattices are
complete with non perturbative renormalization (NPR).  The dynamical results with two degenerate dynamical quark flavors from 1.7 GeV cutoff lattices are without NPR while the axial charge result is naturally renormalized.
\vspace{1pc}
\end{abstract}

\maketitle

\section{INTRODUCTION}

We have been calculating electroweak form factors and moments of structure functions of nucleon using domain wall fermion (DWF) quark action \cite{Kaplan:1992bt,Shamir:1993zy,Narayanan:1992wx,Furman:1994ky} and DBW2 (``doubly blocked Wilson 2'') gauge action \cite{Takaishi:1996xj,deForcrand:1999bi}.  As is well known by now, DWF preserves almost exact chiral symmetry on the lattice by introducing a fictitious fifth dimension in which the symmetry violation is exponentially suppressed.  The DBW2 gauge action,  which contains a rectangular (\(2 \times 1\)) Wilson loop substantially reduces the residual chiral symmetry breaking  compared to the Wilson gauge action at the same size of the fifth dimension~\cite{Aoki:2002vt}.    The combination of domain wall fermions and the DBW2 gauge action has been successfully applied by the RBC group on various aspects of low energy hadron physics where the chiral symmetry plays important roles, such as the mass spectrum and weak matrix elem!
 ents.

We refer ref.\ \cite{Ohta:2003ux} for the details of notations and formulation of the present calculation.  Here we present results obtained from two gauge ensembles \cite{Paris}.  One is quenched with DBW2 gauge action at the lattice cutoff of about \(a^{-1}\) 1.3 GeV.  The other is with two degenerate flavors of dynamical DWF quarks at cutoff of about 1.7 GeV \cite{ChrisTaku} with three sea quark mass values at about 1/2, 3/4 and 1 strange quark mass.  In both cases the residual chiral symmetry breaking is
a few MeV using $L_s=16$ for the quenched case and $L_s=12$ for the dynamical ($L_s$ is the extent of the fifth dimension). There are about 100 configurations for each quark mass values.

\section{AXIAL CHARGE}

We calculate the ratio of the isovector axial and vector charges, \(g_{_{A}}/g_{_{V}}\) which is renormalized by definition.
\begin{figure}
\begin{center}
\includegraphics[width=75mm]{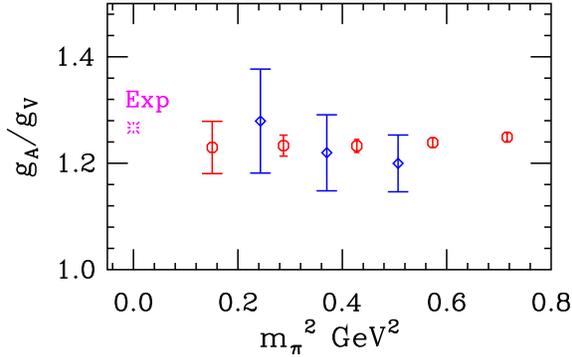}
\end{center}
\caption{Ratio of the isovector axial and vector charges, \(g_{_{A}}/g_{_{V}}\) plotted against pion mass squared.  New \(N_{f}=2\) dynamical results (diamond) seem to follow old quenched (circle) ones and in agreement with experiment (burst at left).  Note weak quark mass dependence.}
\label{fig:gA}
\end{figure}
New this year are the \(N_{f}=2\) dynamical results which are compared with the old quenched ones and the experiment in Fig.\ \ref{fig:gA}.  The dynamical results seem to follow the quenched, with weak dependence on quark mass, and to be in agreement with experiment, though the errors are still large.
  
\section{QUARK DENSITY \(\langle x\rangle_{u-d}\)}

The quenched results are now complete with non-perturbative renormalization (NPR), 
\begin{figure}[t]
\begin{center}
\includegraphics[width=75mm]{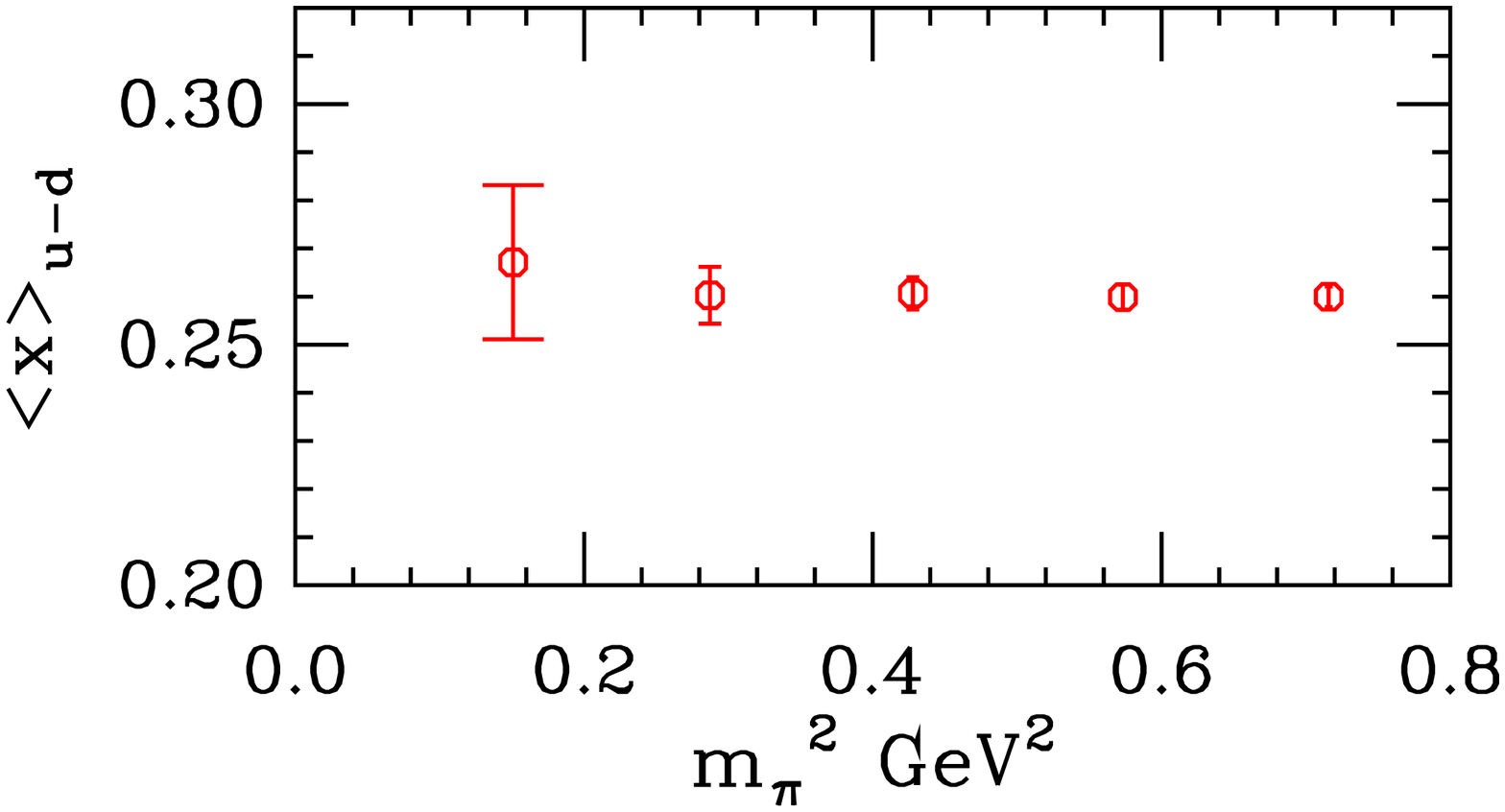}
\includegraphics[width=75mm]{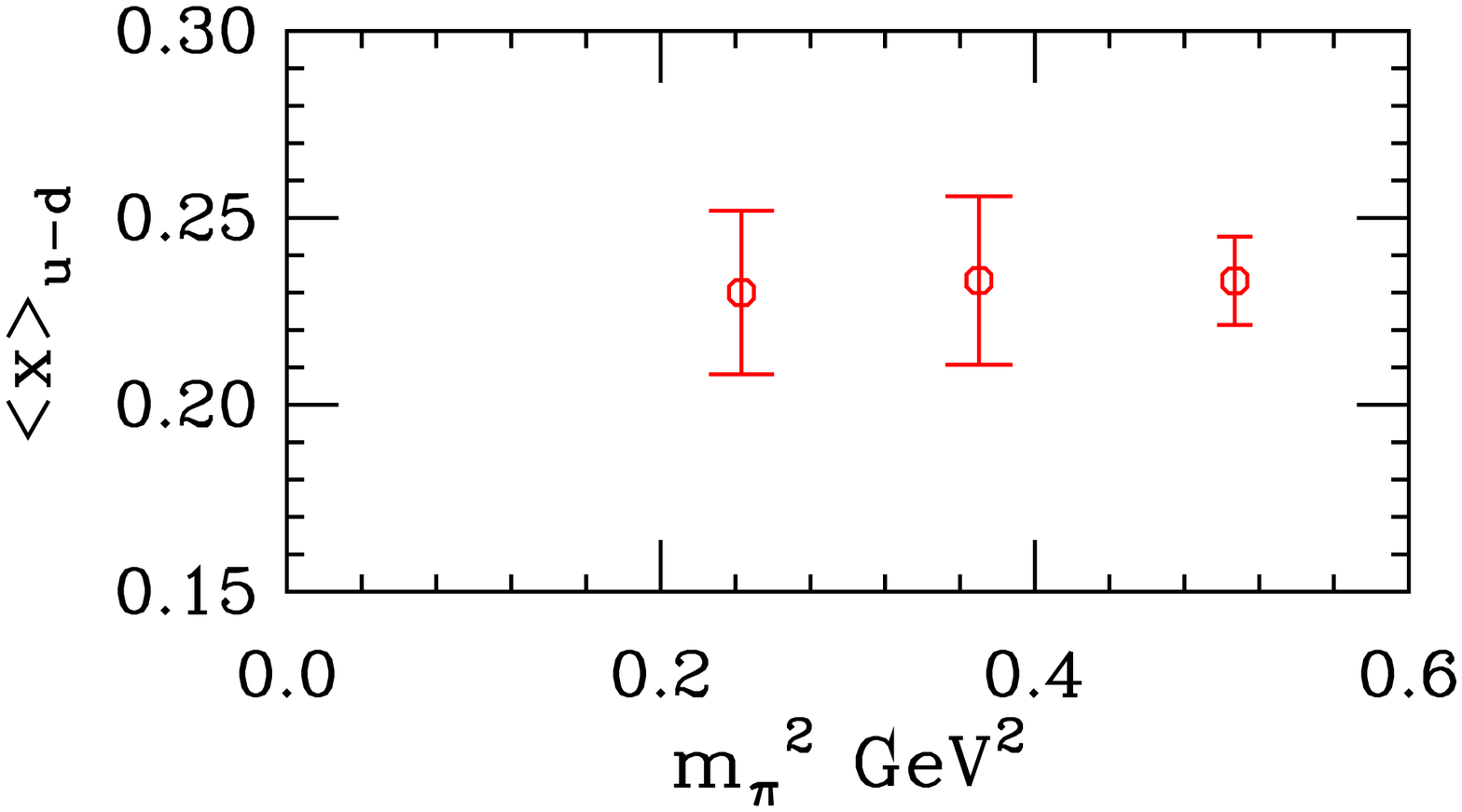}
\end{center}
\caption{Bare quark density, \(\langle x\rangle_{u-d}\).  Quenched results (above).  \(N_{f}=2\) dynamical calculation (below).}
\label{fig:Xns}
\end{figure}
\(Z=1.02(10)\), with \(\overline{\rm MS}\) 2 GeV, 2-loop running.  As is shown in Fig.\ \ref{fig:Xns}, no curvature is seen and the extrapolated value is \(\langle x \rangle_{u}/\langle x \rangle_{d} = 2.41(4)\) at the chiral limit.  Also shown are the ongoing dynamical calculation which lacks NPR and so cannot be directly compared.

\section{HELICITY \(\langle x \rangle_{\Delta u- \Delta d}\)}

Again the quenched calculations are now complete with NPR~\cite{Martinelli:1995ty},
\begin{figure}
\begin{center}
\includegraphics[width=75mm]{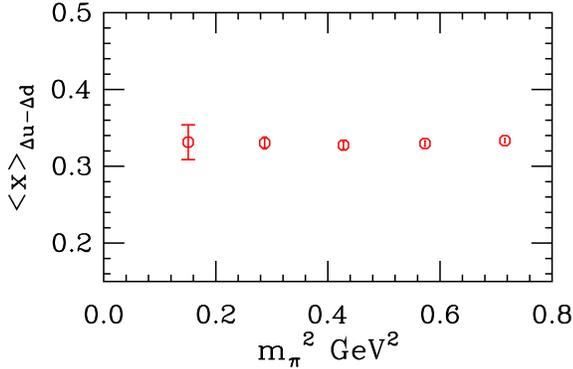}
\end{center}
\caption{Helicity, \(\langle x \rangle_{\Delta u- \Delta d}\).  Bare quenched results.}
\label{fig:XDq}
\end{figure}
\(Z=1.02(9)\), with \(\overline{\rm MS}\) 2 GeV, 2-loop running.  This quantity does not show curvature in the chiral limit either (see Fig.\ \ref{fig:XDq}.)  Corresponding dynamical calculations are ongoing (not shown here).

\section{TRANSVERSITY \(\langle 1 \rangle_{\delta u- \delta d}\)}

Quenched calculations are complete with NPR,
\begin{figure}[t]
\begin{center}
\includegraphics[width=75mm]{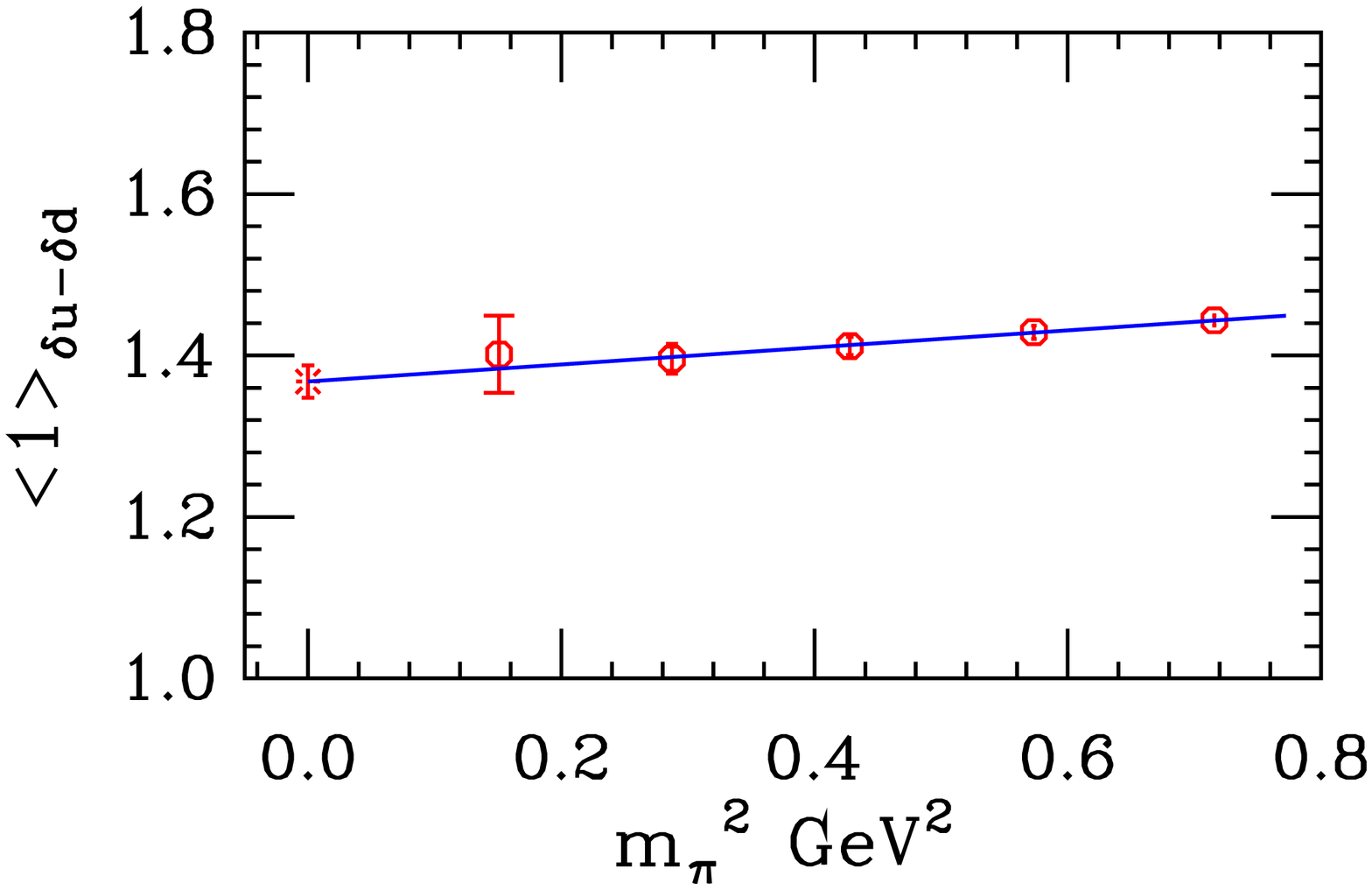}
\includegraphics[width=75mm]{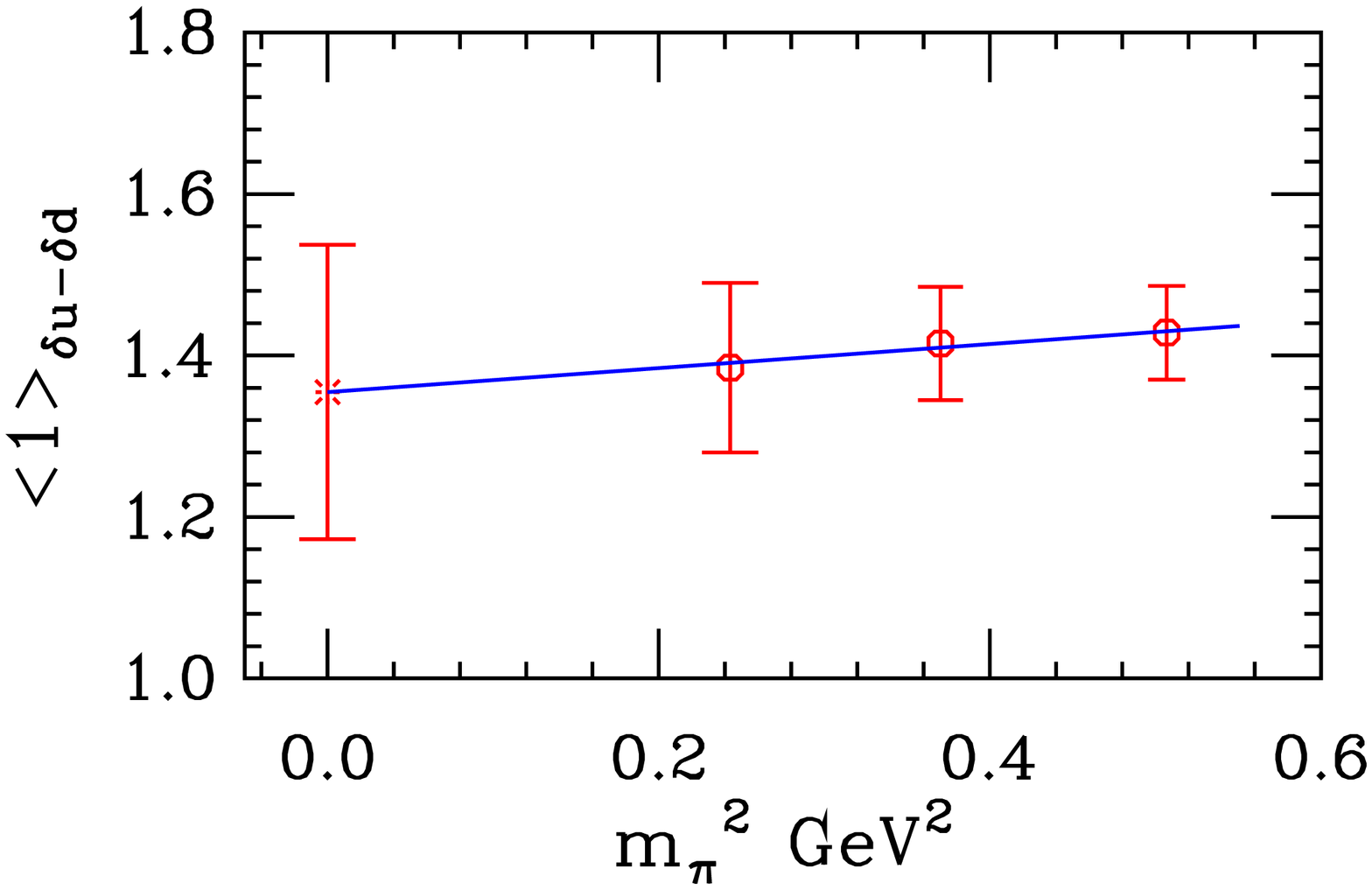}
\end{center}
\caption{Bare transversity,  \(\langle 1 \rangle_{\delta u- \delta d}\).  Quenched results (above).  Dynamical results (below)}
\label{fig:1dq}
\end{figure}
\(\langle 1\rangle_{\delta u-\delta d}=1.193(30)\), \(\overline{\rm MS}\) (2 GeV) 2-loop running.
This result is in agreement with QCDSF quenched continuum result of 1.214(40), \(\overline{\rm MS}\) (2 GeV) showing that the scaling violations with DWF are small for this quantity.  Also shown are the ongoing dynamical calculations, which cannot be directly compared due to lack in NPR.

\section{TWIST 3 \(d_{1}\)}

This is the twist-3 part of \(g_{2}\).  Note that \(\langle x \rangle_{\Delta q}\) is twist-2.
It is negligible in Wandzura-Wilczek relation, \(\displaystyle
g_{2}(x) = -g_{1}(x) +  \int_{x}^{1} \frac{dy}{y} g_{1}(y),\) but need not be small in a confining theory \cite{JJ}.
The quenched calculations, though yet to be renormalized,
appear small in the chiral limit in disagreement with unrenormalized Wilson fermion results~\cite{Dolgov:2002zm}, as we noted before \cite{Ohta:2003ux}.  Note the present calculation does not suffer from power divergent mixing, while the latter does.  We also found the dynamical results, again yet to be renormalized, seem small in the chiral limit.

\section{CONCLUSIONS}

The quenched calculations at \(a^{-1}\) \(\sim\) 1.3 GeV are almost complete with non-perturbative renormalizations (NPRs).  The \(N_{f}=2\) dynamical calculations at \(a^{-1}\) \(\sim\) 1.7 GeV are well under way.     There are many new results:  Dynamical result for the nucleon axial charge, \(g_{_{A}}\), seems to follow the quenched result with weak dependence on the quark mass.  Its chiral extrapolation value is already in agreement with experiment within one standard deviation.  Quenched calculations for moments of structure functions, quark density \(\langle x\rangle_{u-d}\), polarization \(\langle x\rangle_{\Delta u - \Delta d}\), and transversity \(\langle 1\rangle)_{\delta u - \delta d}\), are complete with NPR and do not show appreciative curvature as functions of quark mass down to about 390 MeV pion mass.   Another moment of structure function, \(d_{1}\) seems small in the chiral limit in both quenched and dynamical calculations, but are yet to be renormalized.

In the immediate future, we plan to finish the on-going \(N_{f}=2\) dynamical calculations, perhaps using the new QCDOC computer.  We are also starting to calculate some form factors such as \(F_{1}\), \(F_{2}\), and \(g_{_{P}}\).

\end{document}